\documentclass[draft]{agujournal2019}

\usepackage[T1]{fontenc}
\usepackage[utf8]{inputenc}
\usepackage{textcomp}
\usepackage{url}
\usepackage{lineno}
\usepackage{amsmath}
\usepackage{amssymb}
\usepackage{graphicx}
\usepackage{booktabs}
\usepackage{array}
\usepackage[compatibility=false]{caption}
\usepackage{eurosym}
\usepackage{calc}
\usepackage[inline]{trackchanges}
\usepackage{soul}
\usepackage{placeins}
\usepackage{float}
\usepackage{caption}
\usepackage{multirow}
\usepackage{longtable}
\usepackage{colortbl}
\usepackage{tabularx}
\usepackage{enumitem}
\definecolor{tblhead}{HTML}{D9D9D9}
\definecolor{tblsub}{HTML}{F4F4F4}
\definecolor{tblgrid}{HTML}{808080}

\journalname{Space Weather}

\makeatletter
\def\@oddhead{}
\let\@evenhead\@oddhead
\makeatother

\begin{document}

\title{A State-Level Assessment of Space Weather Impact on the Virginia Power Grid: From Vulnerability to Mitigation Strategies}

\authors{
    Vedaang~Verma$^{1}$,
    Dennies~K.~Bor$^{1}$,
    Lucy~Wilkerson$^{2}$,
    Edward~J.~Oughton$^{1,\ast}$ \\
    \small$^{1}$Department of Geography and Geoinformation Science, George Mason University, Fairfax, VA, USA \\
    \small$^{2}$Department of Physics and Astronomy, George Mason University, Fairfax, VA, USA \\
    \small$^{\ast}$Corresponding author (email: eoughton@gmu.edu)
}


\begin{abstract}
Contemporary methodologies exist for assessing the nationwide impacts of geomagnetically induced currents (GICs), such as the Coupled Space Weather Impact Model (C-SWIM), which links power grid vulnerability with economic losses. Yet regional mitigation requires adapting the national model to the local transmission network so operators and planners can identify substations at risk and evaluate economic and population consequences and protective actions. We present the first adaptation of C-SWIM to Virginia, modeling the Pennsylvania--New Jersey--Maryland (PJM) Interconnection, which includes Dominion Energy Virginia and Appalachian Power. Its extra high voltage network was assessed under the May 2024 Gannon, October 2003 Halloween, and March 2015 St.~Patrick's Day storms in addition to nine return period scenarios derived from statistical fits to 92 historical events. Using the Leontief consumption and production input--output model, we estimated daily losses of \$145.5M for a 100-year storm and \$230.6M for a 250-year storm, with 339{,}000--535{,}000 people and up to 12{,}000 businesses at risk of power outage. Furthermore, the Ghosh input--output model was used to evaluate three island selection algorithms for protective islanding, which temporarily disconnects parts of the grid to reduce transformer damage. Only one algorithm yielded benefits exceeding costs across all twelve scenarios, with avoided losses up to \$9.69M per day and a cost of \$0.204M per intervention. However, these islanding benefits assume disconnection for 6 hours and transformer replacement after 365 days. Moreover, the analysis uses national economic relationships and models only transformer thermal failure, excluding voltage collapse, cascading failures, and restoration. Transmission operators and state energy policymakers can use this statewide assessment to prioritize regional monitoring and appraise protective islanding.

\end{abstract}
\clearpage

\section{Introduction}

Coronal mass ejections (CMEs) from the Sun can trigger geomagnetic storms when directed at Earth. These geomagnetic storms disturb Earth's magnetic field and induce a geoelectric field, whose strength depends on the disturbance and the electrical conductivity of the underlying geology \cite{Love2024}.

The geoelectric field induces voltages along conductors such as power transmission lines \cite{NERC2012GMD,BotelerGIC}, pipelines \cite{Khanal2019,Ingham2025}, railways \cite{Love2019Railroad,Boteler2021,Patterson2024}, and telecommunication systems \cite{Love2025Telecom1921}, driving geomagnetically induced currents (GICs) through circuits connected to ground. In extra high voltage (EHV) networks, transmission lines with low resistance and grounded transformer neutrals provide paths for these quasi-direct currents. GICs can cause half-cycle saturation of transformer cores, increasing reactive power demand, generating harmonics, and causing heating that may damage equipment or contribute to voltage instability and power outages \cite{BotelerGIC,Schachinger2023}.

For example, the March 1989 storm, seen as the archetypal geomagnetic disturbance, collapsed the Hydro-Qu\'ebec grid in about ninety seconds and left over six million people without power during an outage lasting about nine hours \cite{Bolduc2002,Boteler2019,NERC2012GMD}. The 1859 Carrington storm \cite{Carrington1859} predates the modern grid and disrupted telegraph networks worldwide as operators reported sparking equipment \cite{Boteler2006}. A storm of this magnitude hitting the modern highly interconnected EHV power grid could have widespread consequences. As such, historical occurrences necessitate worst-case modeling of geomagnetic storms.

Indeed, \citeA{Oughton2017} estimated daily U.S. losses of \$6.2--41.5 billion across modeled blackout scenarios. On average, more than half of the total losses worldwide occurred outside the blackout zone through links in domestic and international supply chains. More recently, \citeA{oughton2026majorspaceweatherrisks} modeled both direct economic losses from transformer thermal failure and indirect losses through economic supply chains across the United States.

The Coupled Space Weather Impact Model (C-SWIM) uses historical storm data to measure grid vulnerability and estimate the socioeconomic consequences of disruptions to supply and consumption \cite{oughton2026majorspaceweatherrisks}. C-SWIM is applied at the national scale for the United States. While national assessment includes results for individual substations, decision makers, such as transmission operators and state energy offices, operate at regional levels and need to evaluate mitigation on regional networks. For mitigation informed by data to truly be realized, regional figures of economic and population impact need to be derived from storm modeling of the regional network, including specific substations at risk and power grid regions.

Compounding this gap in risk assessment capabilities is a need for the implementation of mitigation informed by data in realistic scenarios. Algorithmic GIC mitigation is a widely studied area. Indeed, a substantial engineering literature demonstrates how to reduce GIC through blocking device placement and corrective control such as transmission line switching \cite{Kazerooni2019,Lu2018,Liang2019,Zhu2015}. However, two limitations remain relevant to regional mitigation. Firstly, mitigation methods are typically evaluated as optimization problems minimizing engineering objectives such as load shedding or power violations. This means that economic loss and the population affected are not necessarily considered when evaluating solutions using engineering criteria alone. Secondly, many mitigation methods are tested on small synthetic test grids unlike the complex networks used in the real world. In actual power grids, interventions such as blocking and switching at any local area can cause unwanted side effects, making local heuristics unreliable \cite{Schachinger2023}. Recent work has begun to close this gap with dedicated benefit-cost analysis of mitigation investment decisions \cite{oughton2026assessingeconomicbenefitsspace}. Mitigation must be decided based on socioeconomic factors and tested via analysis of the entire network on a realistic power grid.

Thus, the following research questions are posited for investigation.
\begin{enumerate}[label=\alph*)]
    \item What are the specific power grid vulnerabilities for the state of Virginia under varying geomagnetic storm scenarios?
    \item How do power grid vulnerabilities translate into economic losses and individuals impacted for the state of Virginia?
    \item How can mitigation informed by data effectively be used to reduce the economic losses incurred due to geomagnetic storms?
\end{enumerate}

This paper attempts to resolve these gaps by presenting the first regional adaptation of C-SWIM for Virginia and a mitigation method of strategic islanding informed by data, justified by cost-benefit analysis. This work's contribution is threefold: (i) a replicable framework for adapting the C-SWIM pipeline that couples physics, engineering, and economics for any local network, producing regional grid vulnerability and economic losses; (ii) actionable findings of geomagnetic storm risk assessment for Virginia; and (iii) a proposed method of protective islanding informed by data through a heuristic algorithm to determine the island and cost-benefit analysis to evaluate the solution.

\section{Literature review}\label{sec:litreview}

This literature review summarizes the current state of peer-reviewed literature on geomagnetic storms, from the physics of GICs to the socioeconomic consequences to mitigation strategies. The literature review was used to justify the materials and methods tested in assessing risk for Virginia and developing solutions informed by data. It also helped determine which methods in the national C-SWIM assessment presented in \citeA{oughton2026majorspaceweatherrisks} to retain or modify for the regional adaptation.

\subsection{GIC physics and transformer vulnerability}

Geomagnetic storms cause variations in Earth's magnetic field that induce a geoelectric field at the surface, with its strength influenced by both the disturbance and the conductivity of the underlying geology \cite{Love2024}. While power grids operate with alternating current (AC) at 50--60 Hz, GICs vary slowly enough to be treated as direct current (DC) in a resistive network, typically at frequencies of 0.1~mHz--0.1~Hz and rarely above 1~Hz. In the framework described by \citeA{BotelerGIC}, the geophysical stage estimates the geoelectric field, and the network stage solves for currents driven by voltages integrated along transmission lines. Grounded transformer neutrals complete the conducting paths.

The network stage can be solved with the Lehtinen-Pirjola method \cite{LPm1985} or a nodal admittance approach, which relates voltages at network nodes to currents entering or leaving them. In networks with multiple voltage levels, the original Lehtinen-Pirjola formulation represents ungrounded nodes through virtual connections with very large grounding resistances. To address this numerical inconvenience, \citeA{LPm2022} proposed the modified Lehtinen-Pirjola (LPm) method, which replaces the earthing impedance matrix with an admittance matrix and represents ungrounded nodes with zero admittance to ground.

Besides computing GICs, risk assessment requires understanding their effects on EHV transformers. GICs bias the magnetic flux in a transformer's core and can cause half-cycle saturation. Locally, stray flux can cause heating in windings and structural components, accelerating insulation aging or causing permanent damage \cite{Dehghanian2021}. Across the network, increased reactive power demand can contribute to voltage instability and collapse \cite{Overbye2012}. Damaged transformers may require replacement. In its 2014 assessment, the U.S. Department of Energy reported procurement lead times of 5--16 months, with longer delays during high demand or supply constraints \cite{doe2014lpt}.

These physical vulnerabilities have been translated into quantitative measurements that can be used to estimate transformer failure probability based on observed GIC. Conventionally, the severity of GIC at any transformer is measured through effective per-phase current, which captures the effects of currents across all windings instead of neutral current alone \cite{Klauber2020AGE}. The North American Electric Reliability Corporation (NERC) specified an effective GIC of 75 A per phase or greater as the threshold for thermal assessment of applicable transformers under the benchmark event in TPL-007-1 \cite{NERC2014}. Beyond a binary threshold, transformer vulnerability can be represented through continuous fragility curves (failure probability as a function of effective per-phase GIC), grounded in mechanisms of insulation aging \cite{Gaunt2016}.

\subsection{Geoelectric hazard characterization}

Constructing a geoelectric hazard model begins with estimating the geomagnetic field. This field is measured directly through a sparse set of ground magnetometers, and spatial interpolation is necessary to reach the estimation density required for GIC modeling. \citeA{Rigler2019} present a comparative analysis of five interpolation techniques against synthetic reference fields generated by a geospace circulation model. They evaluated three techniques based on the input data: nearest neighbor, triangular barycentric, and Gaussian process regression. They also evaluated spherical cap harmonic analysis and spherical elementary current system (SECS) inversion, which incorporate physical information. The authors found that the two methods incorporating physical information generally provided better predictions far from magnetometers, with SECS producing fewer spatial interpolation artifacts.

After interpolating the geomagnetic field, that field must be converted to a geoelectric field, a conversion that is dependent on underlying geology. Magnetotelluric (MT) transfer functions relate the horizontal magnetic field at an MT site to the induced horizontal electric field through tensors that vary with frequency. \citeA{Kelbert2017} describe how MT transfer functions can be combined with geomagnetic time series to estimate local geoelectric fields, and compare their estimates with measurements in Japan. Each transfer function considers subsurface conductivity beneath the MT site, such that a geoelectric field time series for that location can be computed using the local geomagnetic time series and the site's transfer function.

Following the modeling of the geoelectric field, geomagnetic storm scenarios must be constructed for simulation. However, geomagnetic storm risk assessment requires reasoning about severe storms with long return periods that are not captured in a limited observational record of accurate magnetic data. \citeA{Love2018} address this by adapting the extreme-value framework commonly used in natural hazard science to the geoelectric field. Integrating geomagnetic observatory time series with magnetotelluric impedance data, the authors derive statistical distributions of geoelectric amplitude. With these distributions, extreme storm scenarios can be extrapolated and defined by their return period: a 100-year return level is an amplitude expected to be exceeded, on average, once per century under the fitted distribution. \citeA{Love2018} use this principle to construct 100-year geoelectric hazard maps for the Mid-Atlantic United States, while \citeA{Love2019} extend the same methodology across the Northeast, relating the resulting hazard geography to regional geological structure. These hazard patterns are governed by regional crustal geology: electrically resistive igneous and metamorphic rock, such as that of the Appalachian Mountains, produces markedly higher geoelectric amplitudes than the electrically conductive sedimentary rock of the Mid-Atlantic Coastal Plain, with one site in Virginia recording the highest and another the lowest 100-year amplitude of any site across the Northeast \cite{Love2019}. This wide intrastate resistivity gradient makes Virginia a useful case for regional geomagnetic hazard assessment.

\citeA{Lucas2020} extended this approach to geoelectric hazards for the U.S. high-voltage grid. For each MT site, they extracted the peak geoelectric amplitude during each historical storm, and for each transmission line they extracted peak induced voltage, producing a distribution of storm maxima at each location. They fitted lognormal distributions using maximum likelihood estimation to extrapolate return levels. Bootstrapping by resampling the storm maxima and refitting the distributions provided estimates and confidence intervals for those return levels. These intervals describe sampling uncertainty under the fitted statistical model.

While the geophysical stages above are necessary to characterize the hazard, the physical power grid must be modeled for GIC simulations. However, the scarcity of publicly available data on individual grid assets has long constrained space weather risk assessment \cite{Hapgood2016}. \citeA{Bor2026Systematic} address this by proposing a reproducible method for constructing models of transmission networks from open source data. OpenStreetMap (OSM) is used to determine substation locations, while transmission line geometries are derived from the Homeland Infrastructure Foundation-Level Data (HIFLD) \cite{HIFLD}. \citeA{Bor2026Systematic} specifically filter for EHV assets and use spatial integration to map substations to transmission lines. The authors compared modeled ground GICs with measurements from 13 Hall-effect probes in the Tennessee Valley Authority (TVA) region during the May 2024 Gannon storm.

\subsection{Socioeconomic impact modeling of geomagnetic storms}

\citeA{oughton2026majorspaceweatherrisks} assessed the socioeconomic consequences of space weather using C-SWIM at the national scale, coupling physics, engineering, and economics. The authors estimated that a 1-in-250-year geomagnetic storm could cause total U.S. losses of \$1.81 billion per day (95\% confidence interval \$1.65--1.96 billion per day), including direct losses and losses through links between industries. They also estimated that in the same scenario, electricity supply would be lost for about 5.1 million people and 135,000 businesses. However, these estimates cover transformer thermal damage and exclude voltage collapse and cascading grid failures. The authors provide a U.S. benchmark for comparison with regional estimates, subject to differences in network representation and economic assumptions.

Earlier assessments estimated losses for prescribed blackout footprints, while coupled assessments derive the affected area from physical hazard and infrastructure models. \citeA{Oughton2017} estimated U.S. daily losses ranging from \$6.2 billion to \$41.5 billion across four hypothetical blackout scenarios affecting between 8\% and 66\% of the U.S. population. On average, direct losses accounted for approximately 49\% of total losses worldwide, while effects transmitted through international supply chains added \$0.8 to \$7 billion per day to the U.S. losses. It follows that the economic footprint of a geomagnetic storm extends well past the physically affected area. Moreover, the large range of economic figures suggests a sensitivity to assumed blackout extent. In later coupled assessments, the affected region depends on modeled physical hazards and infrastructure vulnerability, although analysts still choose the model assumptions \cite{oughton2026majorspaceweatherrisks}.

\citeA{Eastwood2018} applied an approach using storm scenarios to assess space weather impacts in Europe, parameterizing grid resilience by country and accounting for forecast quality. The results were propagated through value of lost load (VoLL) and multiregional input-output models. For two substorms over western Europe under current forecasting and linear recovery assumptions, the authors estimated direct losses of \euro9.3 billion and global economic costs of \euro787--1,108 billion including international spillovers. The disparity between these estimates illustrates the sensitivity of the economic assessment to the treatment of international spillovers and recovery. The authors used VoLL to value unserved electricity and multiregional input-output analysis to trace losses through economic linkages between regions.

\citeA{Schulte2014} used a detailed multiregional input-output representation of the world economy to model the global consequences of a storm similar to the 1989 event over 187 countries and 15,909 combinations of country and economic sector. Assuming a 10\% loss of electricity supply in the areas of greatest storm intensity and a year for transformer replacement, the authors estimated total global losses of US \$2.4 to \$3.4 trillion over a year, approximately half of these losses coming from countries outside the directly affected zone. These annual losses depend on the assumed outage footprint and duration, and cannot be compared directly with daily loss estimates. Comparisons across studies therefore require consistent geographic coverage, outage severity, and recovery assumptions.

\citeA{Ishii2021} further illustrate the role of geographic scope and outage assumptions in an assessment for Japan. The authors estimated losses of JPY~19.8--23.8 billion for a suspension of electricity transmission throughout the Kanto region lasting 24 hours. They cautioned that stopping electricity supply throughout the region for 24 hours could overestimate losses, and that the assumed event had a very low probability in Japan. Methodological assumptions, the treatment of economic linkages, and geographic scope therefore affect these estimates.

A common analytical framework among these assessments is input-output (IO) modeling, a macroeconomic framework for simulating how critical infrastructure disruption cascades through the economy. Representing the economy through interindustry dependencies, IO modeling translates local loss of electricity supply into indirect losses across sectors based on links through supply chains \cite{GALBUSERA2018186, SUEWING2020104756, Koks2019}. There are two standard formulations: the Leontief model, which relates final demand to production requirements through links to upstream suppliers \cite{MillerBlair2009}, and the Ghosh model, which relates primary inputs to sectoral output through allocation coefficients \cite{Ghosh1958}.

Translating grid vulnerability to socioeconomic consequences requires a spatial link, assigning population and businesses to specific vulnerabilities in the power grid. This is achieved through a method of geographic allocation, in which service territories are constructed around transmission substations via Voronoi tessellation, assigning every location to its nearest substation \cite{Thacker2017}. Population and economic output are apportioned to these territories, allowing analysts to estimate the people and businesses exposed to a substation outage. These territories approximate service areas from proximity and do not establish actual distribution connections. This level of spatial resolution supports assessments of individual substations, which can help transmission operators and state energy offices examine local exposure alongside regional totals.

\subsection{GIC mitigation strategies}

After quantifying the socioeconomic impact of geomagnetic storms, researchers can evaluate GIC mitigation strategies. Geomagnetic disturbance literature commonly focuses on three mitigation strategies: generation redispatch, transmission line switching, and placement of GIC blocking devices \cite{Kazerooni2019}. \citeA{Kazerooni2019} recognize such mitigation strategies as optimization problems, satisfying engineering criteria regarding GIC levels and AC power flow. This framing is similar to much of the following literature, which optimizes corrective controls against engineering metrics. The studies in the following discussion evaluate these controls under engineering constraints, while their objectives and network representations differ in their treatment of regional economic consequences.

A strategy within this optimization paradigm is the work of \citeA{Lu2018}, which integrates transmission line switching, generator redispatch, and load shedding as a mixed-integer nonlinear program. The authors connected storm physics to AC power flow, incorporating transformer thermal limits and reactive power losses as constraints. They used convex relaxations of the AC power flow equations to obtain bounds on the objective and support computation. Two characteristics of this formulation are relevant to regional mitigation. Firstly, the authors minimized generator dispatch and load shedding costs under engineering constraints, without estimating losses across the regional economy. Secondly, the authors evaluated the formulation on test systems. Later robust optimization work also minimized system costs under uncertain geomagnetic disturbances \cite{Ryu2022}. Regional applications therefore require a further link between these operational objectives, the modeled network, and the socioeconomic consequences of outages.

Beyond the choice of objective and test network, local mitigation itself has constraints: since the entire network's conductance structure determines GIC flows, intervention to reduce GIC at one area of the power grid alters current flows elsewhere, introducing unwanted side effects \cite{Schachinger2023}. The spatial reach of these effects is not uniform across scenarios. \citeA{Zhu2015} present a linear sensitivity analysis of blocking at the neutral of a single transformer that found its influence to be largely local. But this locality cannot always be assumed: placement of blocking devices can affect neighboring interconnect systems, meaning that mitigation must be evaluated to avoid relocating the risk as opposed to reducing it \cite{Liang2019}. Therefore, mitigation selected through criteria defined locally must also be accompanied by analysis of the entire network.

The economic benefits of several GIC mitigation strategies were evaluated for New Zealand by \citeA{oughton2026assessingeconomicbenefitsspace}. The authors estimated that a severe storm could cost up to NZ\$3.58 billion in total lost gross domestic product (GDP) over 3 days of national blackout and 3 days of restoration, and that operational strategies including switching and islanding could avoid substantial losses with benefit-cost ratios up to 330 to one. Moreover, GIC mitigation based on research has been put into practice: during the May 2024 "Gannon" storm, the New Zealand network operator implemented a response informed by research, with observations and modeling indicating reduced neutral GIC \cite{MacManus2025}. More broadly, a survey of critical infrastructure operators found that the May 2024 storm prompted 149 enacted mitigation actions across sectors during Solar Cycle 25 \cite{LaNeve2026Mitigation}.

Two facets of this New Zealand application are relevant to the present work. Firstly, engineers and operators implemented targeted line disconnections intended to maintain electricity supply. This operational response differs from deliberate outages within a protected island. Secondly, the North and South Island networks are connected by a high-voltage direct current (HVDC) link, which \citeA{MacManus2025} excluded from their GIC model, treating the two networks separately. Whether islanding is still effective when chosen algorithmically, evaluated against objectives for the entire network, and run on a highly interconnected network such as Virginia's remains to be evaluated.

\subsection{Network science}

A candidate island is a group of substations considered for disconnection from the surrounding grid. The creation and evaluation of these islands can be formulated as a network science problem, in which a power grid is represented as an undirected, weighted graph with substations as nodes and transmission lines as edges. Partitioning a graph into subregions is community detection, for which greedy methods based on modularity are widely used \cite{Clauset2004}. In cases, such as in the present work, where the objective is not a full network partition but the isolation of a single subnetwork, a local search can be applied. A candidate island can be constructed outward from an initial substation by adding neighboring nodes according to a local selection criterion \cite{Clauset2005}. The criterion for accepting a neighboring node determines how such a local search balances the island's internal connections against its boundary.

In determining candidate islands, it is critical to evaluate which substations to include in the isolated subnetwork. Betweenness centrality is a measure of node importance in a graph, introduced by \citeA{Freeman1977} and measuring how often a node lies on the shortest paths between two other nodes. A node with high betweenness lies on many shortest paths, making betweenness a potential proxy for the structural importance of a substation. In power grid applications, however, current distributes across conducting paths according to the circuit resistances, so shortest paths alone do not determine GICs.

This shortcoming is directly addressed by the alternative of current-flow betweenness centrality, introduced independently by \citeA{BrandesFleischer2005} and \citeA{NEWMAN200539}. Instead of measuring betweenness via shortest paths, this method models the network as a resistor circuit and measures passing current through a node as current is injected and withdrawn across a pair of nodes. This is also equivalent to how often a random walk traverses a node. By accounting for all possible paths proportionally by conductance, current-flow betweenness represents current sharing along alternative conducting paths. Yet even this measure depends on how the network is connected (its topology) and on the conductances of those connections, without specifying the geoelectric field of a particular storm. By averaging current through a node across pairs of injection and withdrawal points, the measure does not represent the voltages induced along lines during a particular storm. Therefore, current-flow betweenness centrality fails to account for the direction of geoelectric forcing during a real storm, limiting its use as a proxy for storm exposure.

\clearpage

\section{Materials and methods}

\subsection{Framework overview}

This work adapts C-SWIM, outlined by \citeA{oughton2026majorspaceweatherrisks} and \citeA{Bor2026Systematic}, which combines geophysical hazard modeling, grid vulnerability, and macroeconomic impact assessment, to the state of Virginia and evaluates protective islanding. To produce results for Virginia, the national C-SWIM methodology was largely transferred, with adaptation confined to the network topology and the computations dependent on it. The methods used in the national assessment were retained where they do not depend on the regional network topology.

\begin{figure}[h]
  \centering
  \includegraphics[width=\textwidth,keepaspectratio]{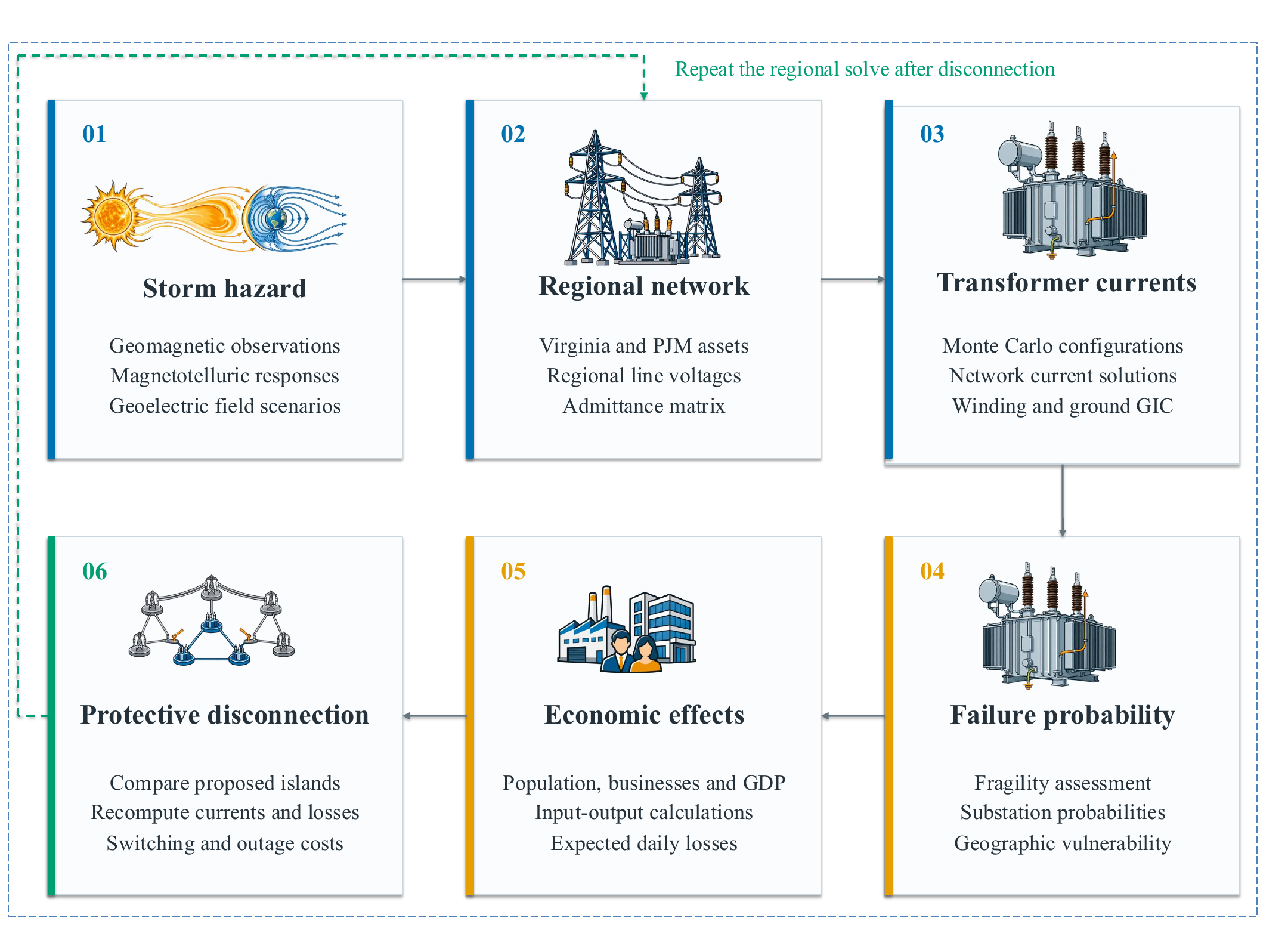}
  \caption{Methodology for adapting C-SWIM to Virginia and evaluating protective islanding.}
  \label{fig:methodology}
\end{figure}

In addition to assessing impacts, the present work proposes mitigation informed by data through protective islanding. This strategy is designed to isolate a portion of the network, temporarily losing power in that island to reduce the risk of damage or failure that would incur major economic losses. Three algorithms were developed to select islands using the estimated population exposure, economic activity, and grid vulnerability. Each algorithm begins at a selected substation and adds neighboring substations according to a cost function. To evaluate islands, the admittance matrix was rebuilt and the GIC simulations repeated, yielding socioeconomic estimates that were compared with estimates without intervention. The full methodology is outlined in Figure~\ref{fig:methodology}.

\subsection{Study domain and network construction}

While impact assessment was focused on the state of Virginia, the constructed electrical model needs to extend past state borders. GICs move through interconnected transmission lines across state borders, meaning that the current flowing through Virginia substations is dependent on the network topology of surrounding states. As such, the study domain included Virginia and surrounding states to represent electrical connections beyond the state border. We approximated the region served by the Pennsylvania--New Jersey--Maryland (PJM) Interconnection using the full state boundaries of Virginia, Maryland, Delaware, New Jersey, Pennsylvania, Ohio, Indiana, Illinois, Michigan, West Virginia, North Carolina, and Kentucky, together with the District of Columbia. This geographic domain differs from the PJM operating footprint, and the effect of extending the electrical network beyond it was not tested. All grid vulnerability results, though evaluated over this regional network, were filtered for Virginia.

The Virginia-PJM power grid was constructed from open source data, following \citeA{Bor2026Systematic}, instead of extracting a portion of the national grid model. This ensures that the electrical model is constructed with a documentable, transferable process, supporting the goal of a reproducible framework for analysis of any power grid region. EHV transmission lines were defined to have nominal voltages of at least 230~kV and were derived from HIFLD, spatially filtered to the study domain. EHV substation locations were determined through OSM via the Overpass application programming interface (API) \cite{Olbricht2015Overpass}, and transmission lines were linked to substations where their geometries intersected a 150~m buffer around a substation. The resulting network contained a main connected component and smaller disconnected components. Each substation was assigned a unique identifier, while the length of each transmission line was computed from its geometry. Its resistance was calculated from that length and a resistance per unit length assigned by voltage level.

\begin{figure}[h]
  \centering
  \includegraphics[width=\textwidth,keepaspectratio]{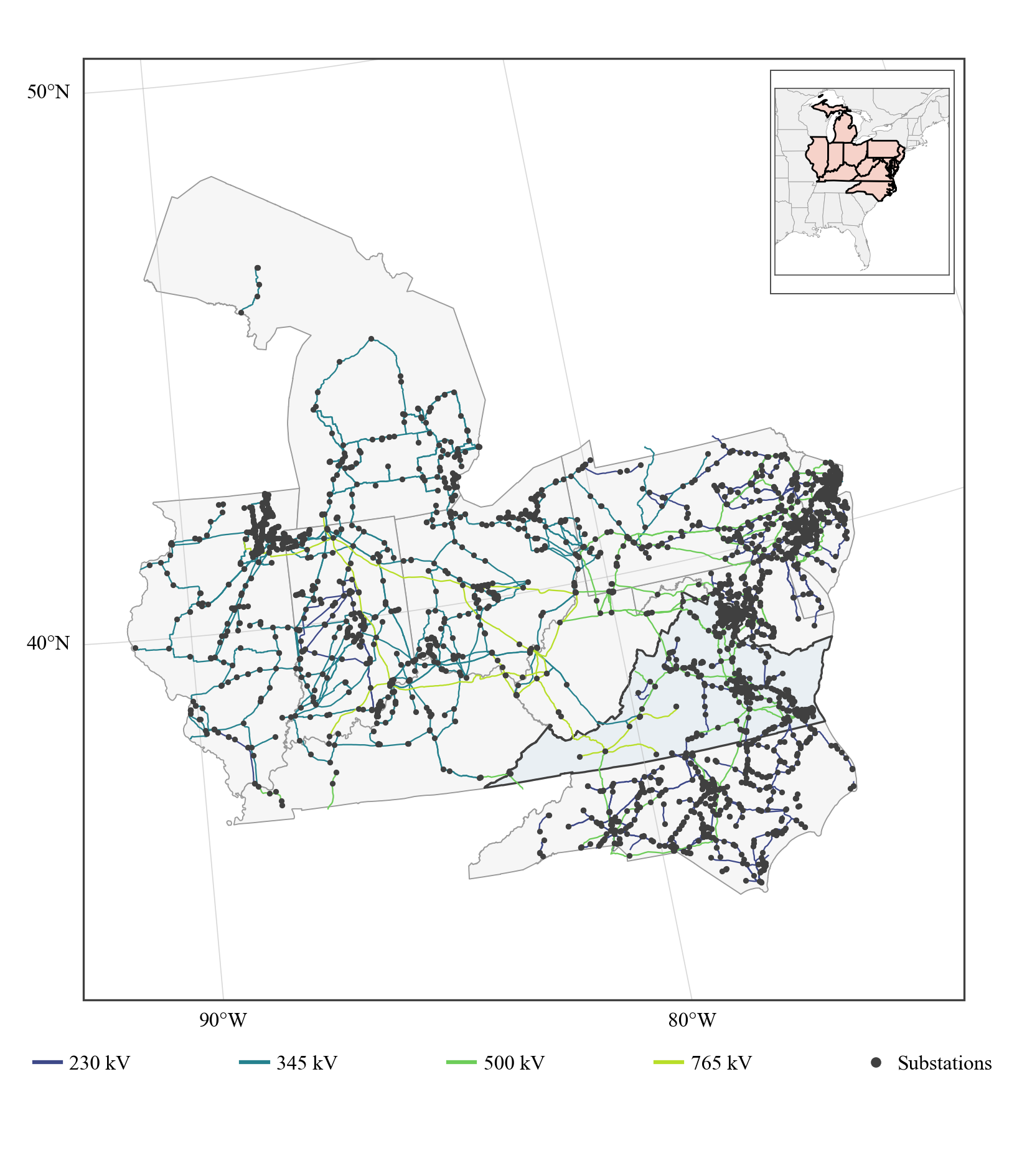}
  \caption{Transmission network for Virginia and the surrounding study domain, constructed from OSM and HIFLD data.}
  \label{fig:VAnetwork}
\end{figure}

The resulting network comprises 2,624 substations and 4,586 transmission lines, as shown in Figure~\ref{fig:VAnetwork}.

\subsection{Geoelectric hazard characterization}

The geoelectric field induced at Earth’s surface was characterized using the 92 historical storms in the national C-SWIM assessment. We estimated the geomagnetic field using SECS interpolation \cite{Rigler2019} of ground magnetometer data from USGS and NRCan geomagnetic observatories, retrieved via the INTERMAGNET data service for storms from 1991 onward and directly from each agency’s own archive for earlier storms \cite{USGSGeomag,NRCanGeomag,INTERMAGNET}, then derived the geoelectric field by convolution with MT transfer functions, following \citeA{Lucas2020}. Transfer functions from the EarthScope/USArray survey were obtained from the EarthScope Magnetotelluric Transfer Function database \cite{Kelbert2011,Kelbert2017}. Critically, these calculations depend on the geomagnetic observations and local conductivity response but are independent of the power grid topology, so the field calculations from the national assessment were retained.

All 1,696 MT sites from the national calculation were retained to construct the spatial interpolation weights for integrating the field along regional transmission lines. Sites outside Virginia therefore contributed to the interpolation along lines crossing the study domain.

Storm scenarios in this work were constructed and classified by return period. For each site and transmission line in the study domain, we fitted lognormal distributions to the magnitudes of storm maxima from the 92 historical storms. The fitted quantities were the geoelectric field at each site and the induced voltage along each transmission line. Nine return periods from 50 to 250 years were evaluated in increments of 25 years. Each synthetic scenario consists of estimates fitted separately at individual sites or lines and does not represent the evolution of the field across the region during a single storm. For comparison with these synthetic storms, named historical storms are used: the May 2024 Gannon storm \cite{Wilkerson2026GIC}, the October 2003 Halloween storm, and the March 2015 St.~Patrick's Day storm.

We used the three historical storms to evaluate protective islanding under different patterns of positive and negative voltages along transmission lines. While voltage magnitudes for the synthetic scenarios were fitted independently for each line, the voltage signs follow a common pattern derived from the Gannon storm. Thus, the nine return periods test different voltage magnitudes under the same pattern of voltage signs, and the historical cases test other patterns of voltage signs along the transmission lines.

\subsection{Line voltage recomputation}

Induced line voltages, however, are critically dependent on grid topology. Following \citeA{BotelerGIC}, the voltage induced along a transmission line is the integral of the geoelectric field along that line's specific path, depending on a line's length and orientation. These voltages depend on the paths of the transmission lines in the modeled network. As such, the induced voltages for the Virginia-PJM power grid cannot be obtained by selecting values from the national results.

Consequently, induced voltages were recomputed for the Virginia-PJM network by integrating the geoelectric field from the national calculation along each newly modeled transmission line across all storm scenarios. For regional adaptations of a national GIC assessment, the geoelectric field calculations can be retained, while induced voltages must be recomputed for the regional transmission lines. Making this distinction is critical in a regional adaptation to avoid unnecessary recomputation and invalid reuse.

\subsection{GIC calculation}

After the network topology was constructed and the induced line voltages computed, GICs were computed by solving the network as a resistive DC circuit. The nodal admittance (Y) matrix used line resistances and substation grounding to represent the power grid topology of the Virginia-PJM region. This followed the LPm formulation of \citeA{LPm2022}. Given the induced line voltages, we solved for nodal voltages, branch currents, and neutral-to-ground currents.

However, a single GIC calculation does not capture uncertainty in transformer configurations, which are not publicly available at the substation level \cite{Bor2026Systematic}. Following the national assessment, transformer types (autotransformer, wye-wye, and wye-delta) and the number of transformer units per substation were selected at random for each of 3,000 Monte Carlo samples. GICs were calculated for each sample to obtain a distribution of currents at each substation. We used the mean and 95th percentile to summarize current exposure, while the fragility calculations used the GICs calculated for each sample.

To measure GIC, both winding and neutral-to-ground currents were computed. Winding GIC per phase was used for the substation ranking. Transformer heating depends on the currents in individual windings and the transformer configuration \cite{Klauber2020AGE}.

\subsection{Vulnerability and economic analysis}

The GICs calculated for each sample were converted into estimated failure probabilities using continuous fragility curves. For these calculations, we used $I_E=|I_g|/3$, where $I_g$ is the neutral-to-ground current at a substation, as a proxy for current per phase. This proxy differs from the winding current used in the substation ranking and does not resolve the thermal response of each transformer configuration. The NERC TPL-007-1 value of 75~A/phase is a threshold for thermal assessment, not an empirical threshold for transformer failure \cite{NERC2014}. Specifically, the assumed failure probability for the current proxy $I_E$ is
\begin{equation}
P_f(I_E) = \Phi\!\left(\frac{\ln I_E - \mu}{\sigma}\right),
\label{eq:fragility}
\end{equation}
where $\Phi$ is the standard normal cumulative distribution function and $\mu$ and $\sigma$ are the fragility parameters. The fragility calculation uses $\mu=\ln(\theta/C)$, with a current scale that depends on age, $\theta=75[1-0.4F_{\mathrm{age}}]$~A/phase and a sampled current scaling factor $C$ between 0.60 and 1.00. Here, $F_{\mathrm{age}}$ is a Weibull cumulative aging factor, and $\sigma$ is sampled uniformly between 0.25 and 0.50. For each Monte Carlo sample and storm scenario, aging and fragility parameters were repeatedly selected at random, and failure outcomes were sampled using the resulting probabilities. We retained these assumptions without calibrating them to individual Virginia transformers. The continuous curve represents incremental changes in assumed failure probability as GIC changes. Figure~\ref{fig:transformerfailure} summarizes failure probabilities for the first 2,000 of the 3,000 Monte Carlo samples, while the economic estimates use all 3,000.

In order to derive a storm's economic and human footprint from substation failure, a spatial link between grid regions and their surrounding economic activity is necessary. Following \citeA{Thacker2017} and the national assessment \cite{oughton2026majorspaceweatherrisks}, approximate service areas were constructed around each substation using Voronoi tessellation based on proximity. Derived from 2020 ZIP Code Tabulation Area (ZCTA) boundaries \cite{USCensusZCTA2020} and 2020 decennial census population counts \cite{USCensus2020PL94171}, economic activity, population, and establishments were then allocated within these areas using dasymetric interpolation with National Land Cover Database (NLCD) land cover \cite{USGSAnnualNLCD2024}. In this manner, activity was apportioned using a land cover mask, and the allocated values were rescaled to preserve the totals in the source ZCTAs. As such, a substation's assumed failure was linked to the population and establishments assigned to its approximate service area.

In order to assess impact across the economy, localized losses are translated into economic consequences for each sector through three techniques of IO modeling. A Ghosh model for reductions in value added, a Leontief model for reductions in consumption, and a Leontief model combining consumption and production effects were all used. These approaches compare different assumptions about how an electricity disruption affects economic activity.

The methodology for economic analysis does have a scope limitation. The IO analysis in this work uses national production and final demand tables, to which reductions in value added and final demand associated with outages in Virginia were applied. We did not construct a social accounting matrix (SAM) or IO tables for Virginia as that was out of the scope of this present work. The estimated indirect losses therefore depend on national relationships among sectors and do not establish where those losses occur geographically. For the economic sensitivity ranges, the 5th and 95th percentiles of the affected population and of affected GDP in each sector were applied separately to the IO calculations. These ranges describe sensitivity to the resulting combinations of economic inputs and are not joint confidence intervals for total loss.

\subsection{Protective islanding}

An algorithmic approach to protective islanding was implemented in this work for GIC mitigation. Protective islanding intentionally isolates a portion of the network, leaving it without power for a short duration to reduce the risk of transformer failure. The network is reconnected after this short period, potentially avoiding the long lead times of transformer replacement.

This work evaluated three original algorithms for island selection, yielding three distinct islands that were assessed through cost-benefit analysis. The algorithms are outlined in Figure~\ref{fig:islanding_methods}, with symbols defined in Table~\ref{tab:islanding-legends} in the appendix.

\subsubsection{Island A: Betweenness centrality}

The selection algorithm for Island A is characterized by its use of betweenness centrality from Section~\ref{sec:litreview} in the formulation of the Substation Criticality Index (SCI), which is an indicator of a substation's relative importance. The greedy algorithm begins island formation at the substation with the greatest SCI and repeatedly appends a neighboring substation until no neighboring substation increases the ratio between the island's aggregate SCI and a cost score for its boundary lines. This search score sums the nominal line voltages divided by 100~kV and is distinct from the monetary switching cost used in the economic evaluation.

\subsubsection{Island B: Current-flow betweenness and redistribution}

The selection algorithm for Island B addressed two deficits in the selection algorithm for Island A. Firstly, current-flow betweenness was used instead of betweenness centrality as a better proxy for how current flows through a graph. Secondly, a measure of redistribution was introduced since intervention at one portion of a network can impact the surrounding network, as established in Section~\ref{sec:litreview}. 

\subsubsection{Island C: Current-flow betweenness and admittance calculations for historical storms}
The selection algorithm for Island C addresses a key deficit of the selection algorithms of Island A and Island B. During a real storm, the directions of the induced line voltages affect GICs, while betweenness centrality and current-flow betweenness do not account for those directions. For Island C, the admittance equations were solved under the three historical storms to identify candidate substations and evaluate islands. During the search, candidates were evaluated using one Monte Carlo sample, and a candidate was retained only when its estimated change in loss was favorable across all three historical cases. The selected island was then evaluated using all of the Monte Carlo samples for the economic comparison.

\subsubsection{Economic evaluation}

Economic evaluation for each island focused on the difference in estimated losses with and without islanding. GICs and estimated failure probabilities were calculated for both network configurations, followed by separate estimates of economic losses. This was repeated across synthetic return periods and historical storms and yielded a difference in terms of GDP loss per day, enabling cost-benefit analysis.

Key assumptions were made for the economic evaluation. Firstly, a disconnection period of 6 hours was assumed to estimate the cost of the temporary outage, informed by the duration of a geomagnetic storm's main phase, which typically lasts several hours \cite{gonzalez1994geomagnetic}. The comparison used static network topologies and did not simulate the timing of disconnection or reconnection within a storm.

To translate daily economic losses into a single economic benefit estimate, damaged transformer replacement times needed to be considered. The U.S. Department of Energy reported procurement lead times of 5--16 months in 2014, with longer delays under high demand or supply constraints \cite{doe2014lpt}. Replacement duration was treated as a scenario assumption, with alternatives of 180, 365, and 730 days. The results in this work are reported for a replacement time of 365 days, so the estimated benefits are conditional on that duration.

To estimate cost, a switching cost of \$25,000 per boundary line was assumed for each of two operations, opening the line before the storm and reconnecting it afterward, an order of magnitude below the total switching-cost assumption used in a comparable New Zealand islanding mitigation study \cite{oughton2026assessingeconomicbenefitsspace} and scaled down since that study’s figure covered a national multi-line plan rather than a single island’s boundary. This cost was an assumed planning value and was not obtained from utility operating records.

The economic benefit was calculated by multiplying the difference in daily GDP loss with and without islanding by the assumed transformer replacement duration. These daily losses were estimated with the Ghosh input-output model from the estimated failure probabilities. The cost included the temporary outage for the island and any substations outside it that lost their connection to the surrounding grid when the island was disconnected, together with switching cost. The outage cost was calculated by prorating each affected substation's allocated annual GDP to the 6-hour disconnection window, as the allocated annual GDP multiplied by $6/(365\times24)$, the fraction of a year that 6 hours represents. The benefit-cost ratios were calculated as the quotient of economic benefit and cost.

\begin{figure}[!ht]
  \includegraphics[width=\textwidth,keepaspectratio]{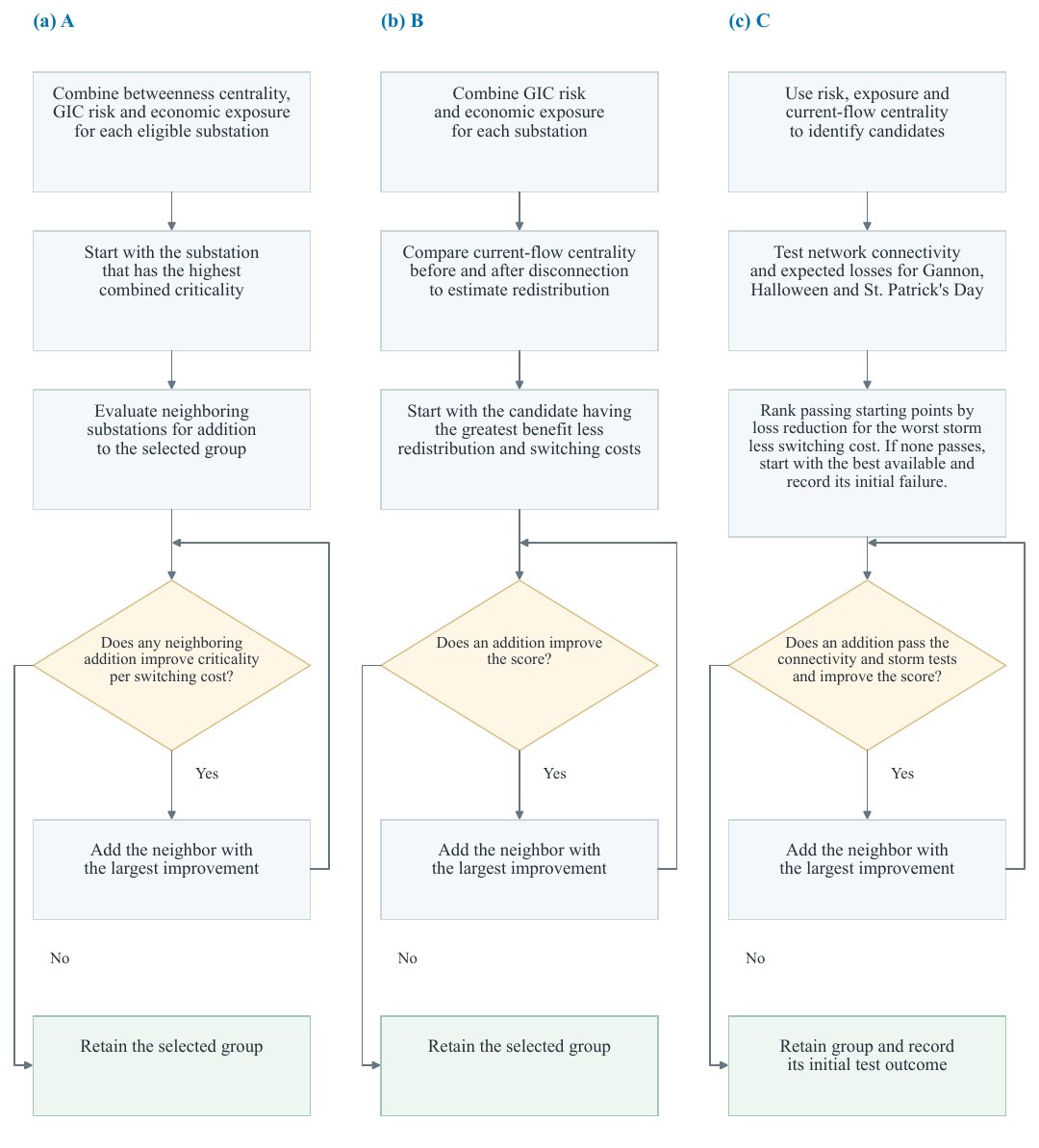}
  \caption{Flowcharts for island selection algorithms}
  \label{fig:islanding_methods}
\end{figure}
\clearpage
\section{Results}

\begin{figure}[!ht]
  \centering
  \includegraphics[width=1.1 \textwidth,keepaspectratio]{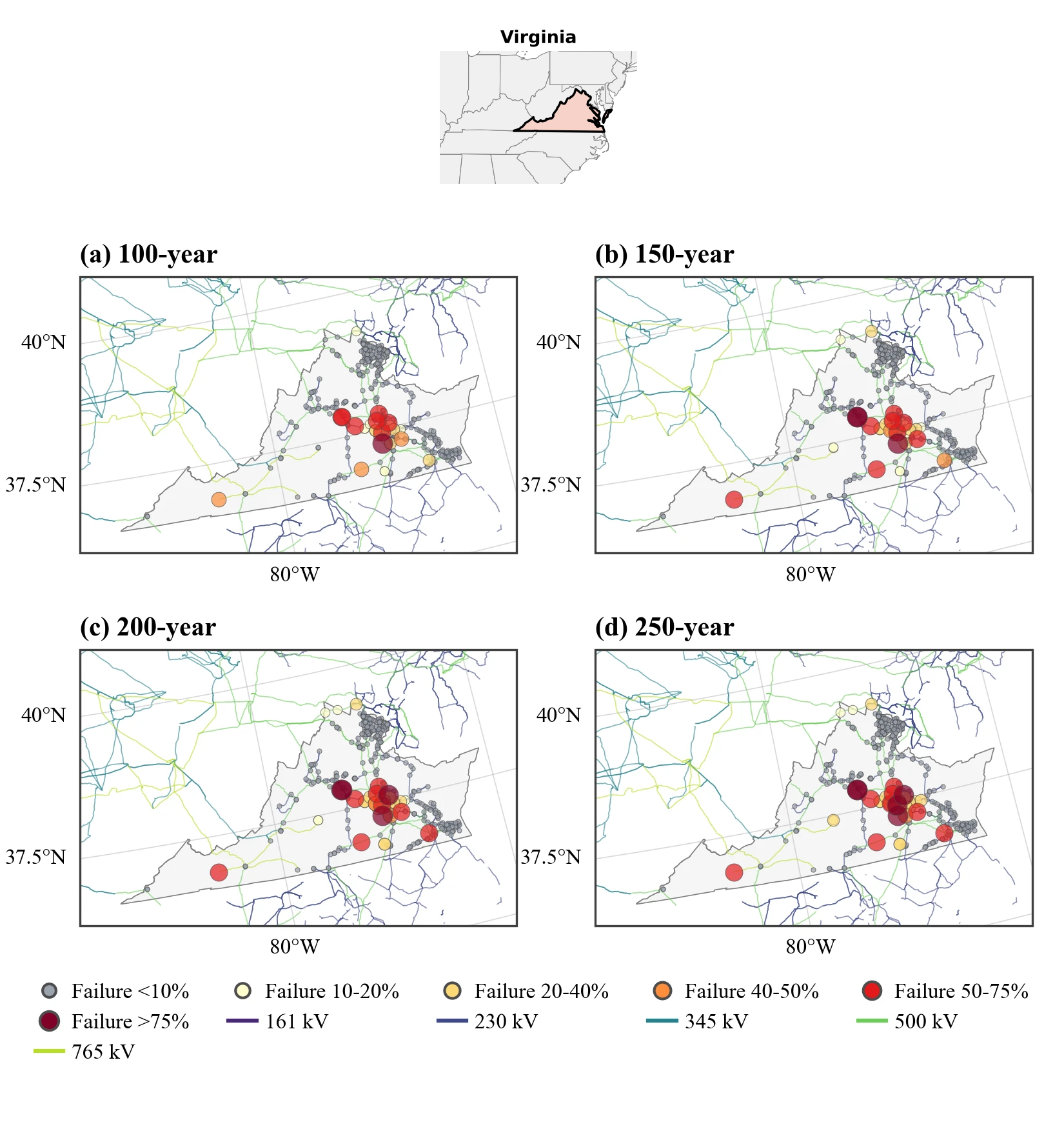}
  \caption{Mean modeled transformer failure probability at Virginia substations for four return periods, using 2,000 Monte Carlo samples.}
  \label{fig:transformerfailure}
\end{figure}

\begin{figure}[!ht]
  \centering
  \includegraphics[width=\textwidth,keepaspectratio]{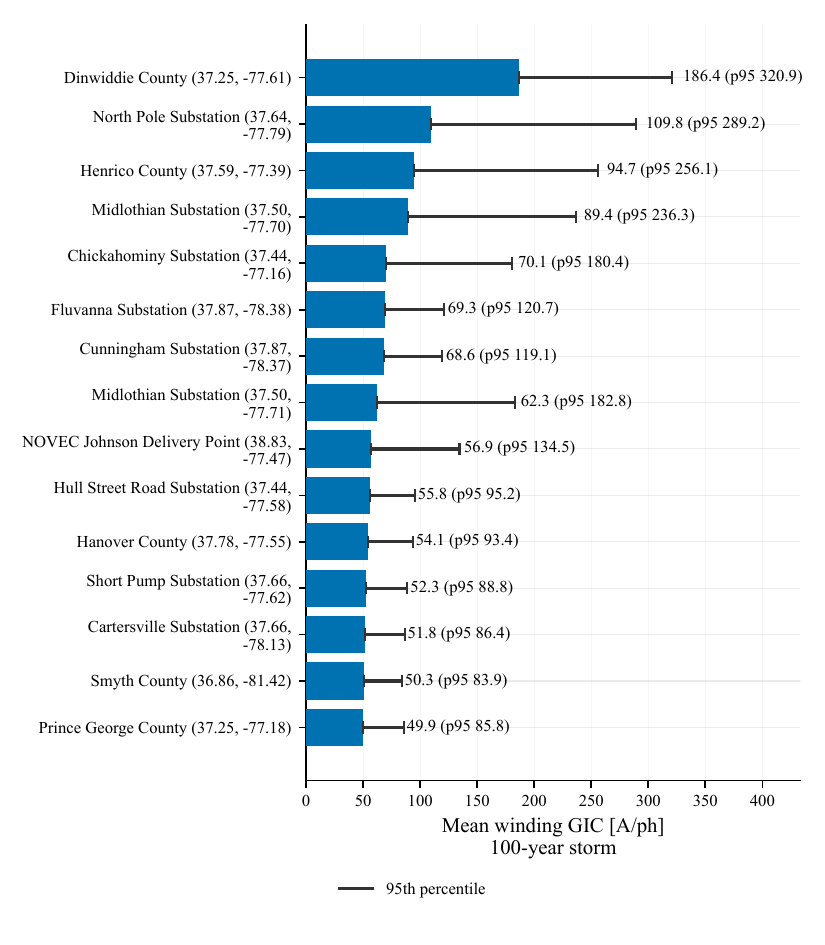}
  \caption{Fifteen Virginia substations with the largest mean absolute winding GIC for the 1-in-100-years scenario.}
  \label{fig:subsranking}
\end{figure}

\begin{figure}[h]
  \centering
  \includegraphics[width=\textwidth,keepaspectratio]{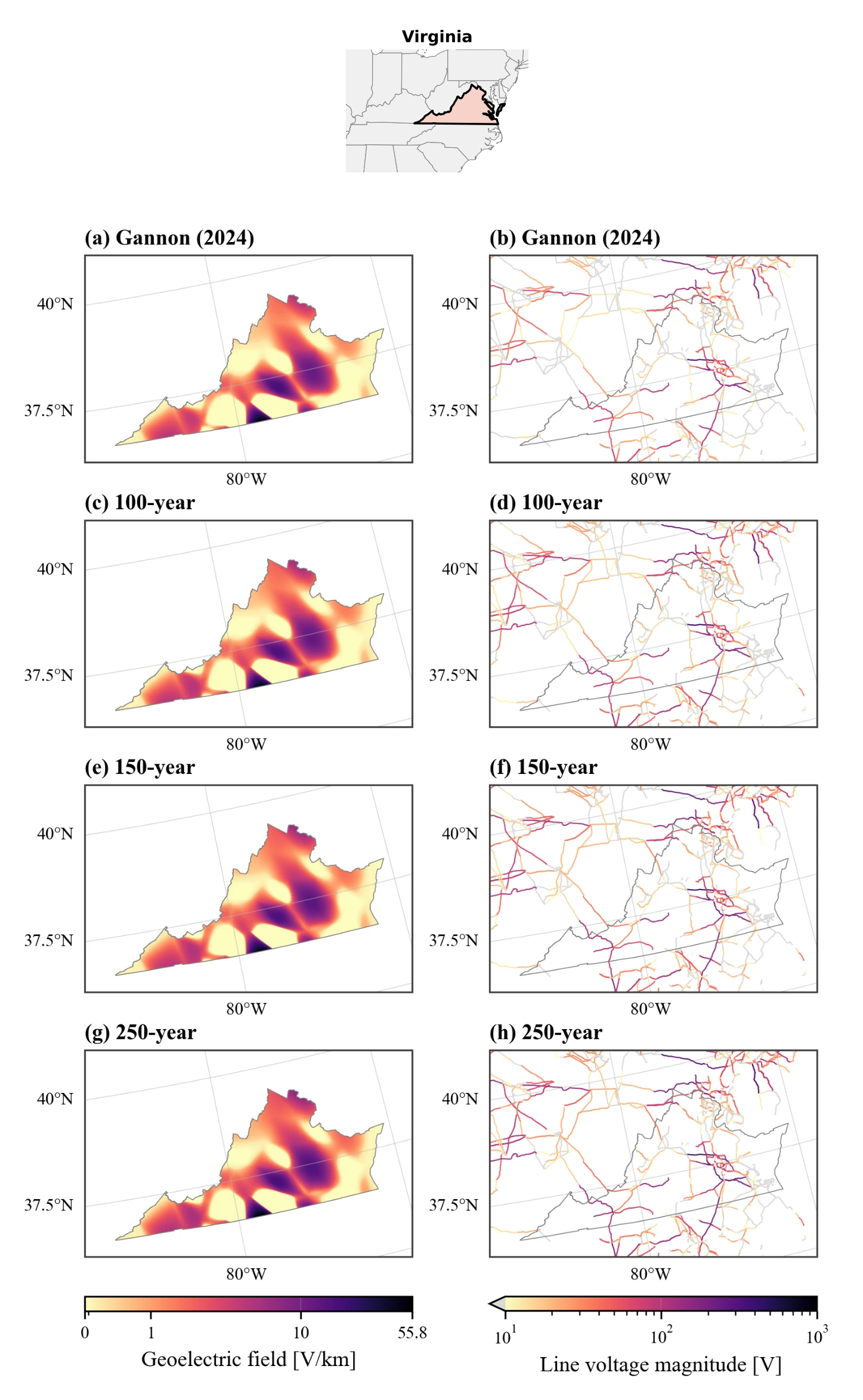}
  \caption{Geoelectric field and transmission line voltage magnitudes for the Gannon storm and selected return periods.}
  \label{fig:hazardmap}
\end{figure}
We calculated GICs for individual transformer windings and evaluated failure probability at each substation, allowing us to identify locations for further mitigation assessment. Figure~\ref{fig:transformerfailure} shows the mean modeled transformer failure probability at each substation for different return periods, using one third of the absolute neutral current as a proxy for current per phase in the lognormal fragility calculation. The results reveal that estimated risk is heavily concentrated in central Virginia, particularly in the Richmond region ($37.54^{\circ}$ N, $77.44^{\circ}$ W). This is consistent with a relatively large estimated geoelectric field amplitude for the region, as shown in Figure~\ref{fig:hazardmap}, which is shaped in part by Virginia's underlying geology: the highly resistive Appalachian basement rock that underlies much of the state's western half tends to produce a comparatively stronger geoelectric field for a given geomagnetic disturbance. This also coincides with the intersection of many high-voltage corridors in the Richmond area, as shown in Figure~\ref{fig:VAnetwork}.

A ranking of Virginia substations based on absolute winding GIC is shown in Figure~\ref{fig:subsranking}. The substation in Dinwiddie County at $37.25^{\circ}$ N, $77.61^{\circ}$ W had the largest mean winding GIC, and many substations with large currents are located near the Richmond area. Transmission operators and state energy officials can use this ranking to identify substations for further monitoring and mitigation assessment, after confirming equipment identities and operating conditions.
\begin{figure}[h]
  \centering
  \includegraphics[width=\textwidth,keepaspectratio]{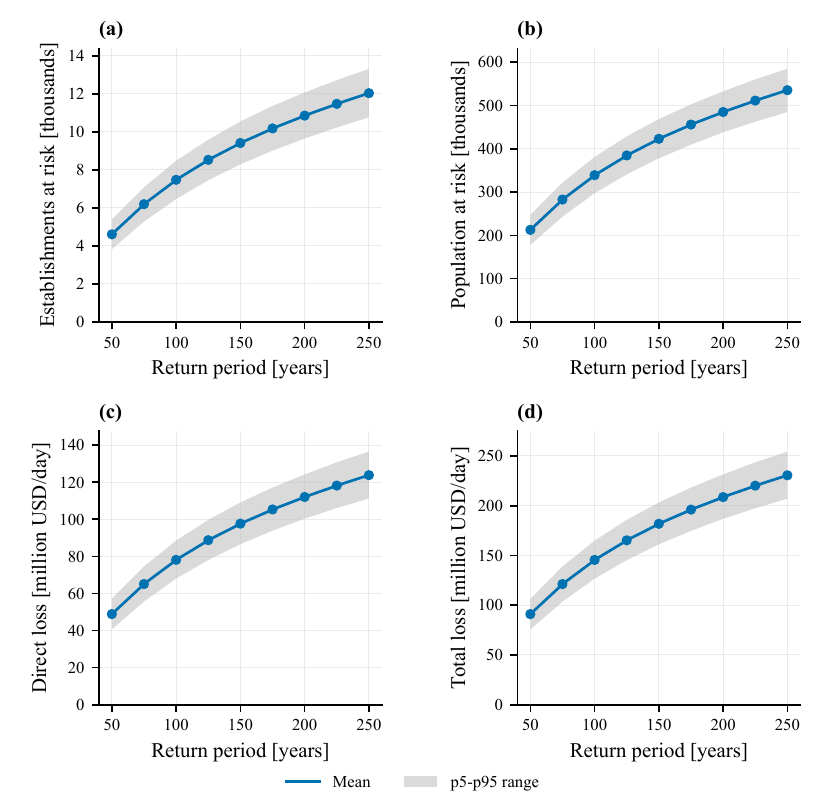}
  \caption{Estimated population, establishments, and daily economic losses by return period using the Leontief consumption and production model.}
  \label{fig:severityrisk}
\end{figure}
\begin{table}[h]
\centering
\caption{Estimated daily economic and human impact on Virginia by storm
return period, and for the 2024 Gannon storm (Leontief consumption and
production model). Economic losses are reported as means [lower, upper],
with ranges evaluated using the 5th and 95th percentiles of the affected population and of affected GDP in each sector.}
\label{tab:econ_impact}
\resizebox{\textwidth}{!}{%
\begin{tabular}{lcccc}
\toprule
\rowcolor{tblhead}
\textbf{Return period} & \textbf{Total loss (\$M/day)} & \textbf{Direct loss (\$M/day)} & \textbf{Pop.\ at risk (k)} & \textbf{Establishments at risk (k)} \\
\midrule
100-year      & $145.5\ [126.7, 164.8]$ & $78.1\ [68.0, 88.5]$ & 339 & 7.5 \\
150-year      & $181.8\ [161.2, 203.3]$ & $97.6\ [86.6, 109.1]$ & 423 & 9.4 \\
200-year      & $208.7\ [186.9, 231.4]$ & $112.0\ [100.4, 124.2]$ & 485 & 10.9 \\
250-year      & $230.6\ [207.1, 254.4]$ & $123.8\ [111.2, 136.5]$ & 535 & 12.0 \\
Gannon (2024) & $137.8\ [121.2, 153.4]$ & $73.8\ [65.0, 82.2]$ & 323 & 6.4 \\
\bottomrule
\end{tabular}%
}
\end{table}

\begin{figure}[!ht]
  \centering
  \includegraphics[width=\textwidth,keepaspectratio]{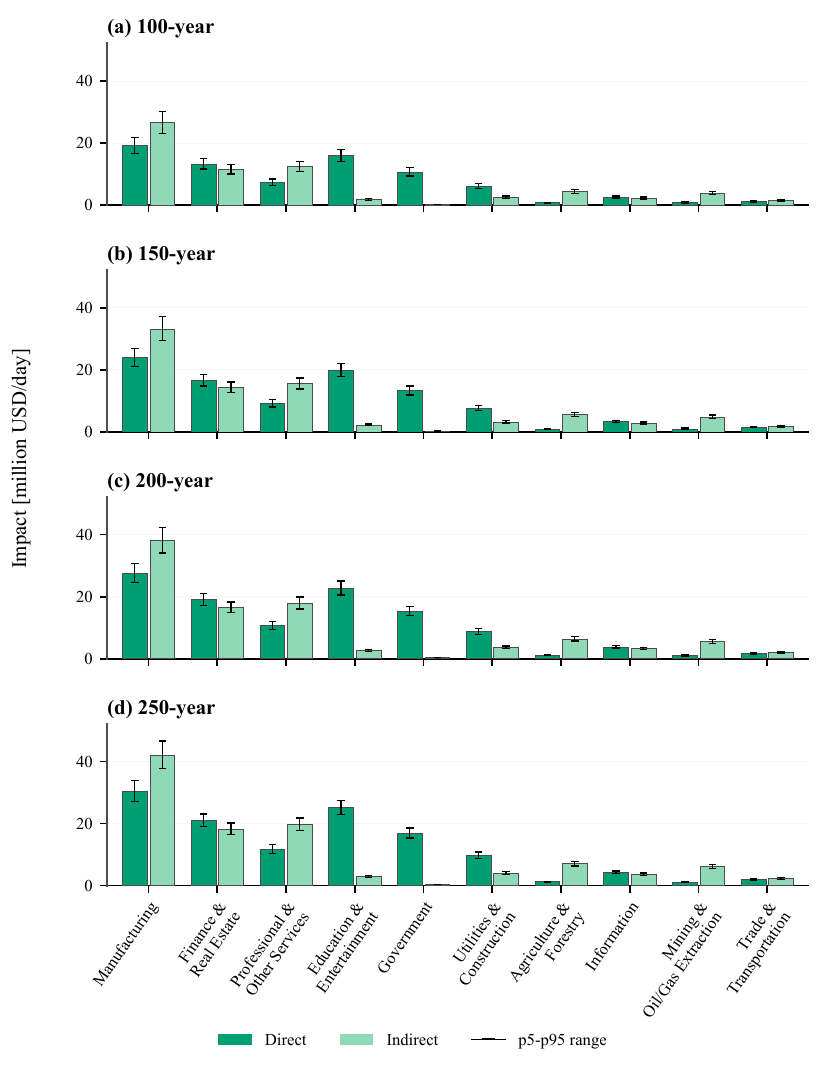}
  \caption{Estimated daily direct and indirect economic losses by sector under the Leontief consumption and production model.}
  \label{fig:LeontiefCP}
\end{figure}

Beyond physical vulnerability, socioeconomic impacts were estimated through input-output modeling. Figure~\ref{fig:severityrisk} and Table~\ref{tab:econ_impact} report daily economic losses and human impact under the Leontief consumption and production model for varying return periods. Using the Leontief consumption and production model, we estimated total daily losses of \$145.5M for the 100-year scenario and \$230.6M for the 250-year scenario, with 339,000--535,000 people and up to 12,000 businesses at risk for the latter. Estimated daily losses were greater for longer return periods, but differences between successive periods were smaller at the longer return periods evaluated. Notably, daily losses were estimated at \$137.8M using the May 2024 Gannon storm drivers, slightly below the estimate for the 100-year scenario. Critically, indirect losses between industries accounted for an average of ~46\% of total daily losses across tested return periods.

Sector-level losses were computed using the same ten NAICS-based aggregate sector categories and national input-output structure as the national C-SWIM assessment \cite{oughton2026majorspaceweatherrisks}, applied to Virginia-specific reductions in value added and final demand derived from this work's substation-level failure and exposure analysis. The sector categories and inter-industry linkages are therefore national, not re-derived for Virginia, consistent with this work not constructing a Virginia-specific social accounting matrix; only the magnitude and location of the underlying shock are specific to Virginia. This explains why the relative ranking of sectors closely mirrors the national assessment. Manufacturing and finance, insurance, and real estate had the largest estimated losses, as shown in Figure~\ref{fig:LeontiefCP}, consistent with findings from the national assessment \cite{oughton2026majorspaceweatherrisks}. Manufacturing alone accounts for the largest share, driven predominantly by indirect effects transmitted through supply chains.

For context, the estimated daily loss associated with the Virginia disruption under a 250-year storm is approximately 13\% of the published national estimate of \$1.81B per day for the same return period \cite{oughton2026majorspaceweatherrisks}. This comparison is approximate because the regional and national assessments use different network models and input-output formulations.

\begin{figure}[!ht]
  \includegraphics[width=\textwidth,keepaspectratio]{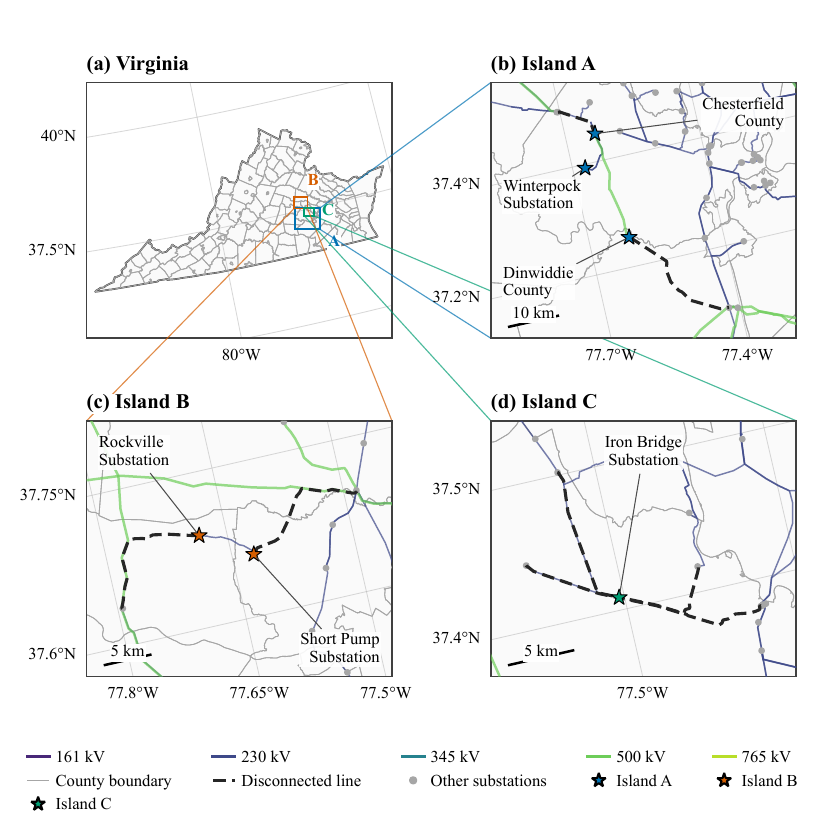}
  \caption{Locations and transmission connections of Islands A, B, and C within Virginia with substations included in each island labeled.}
  \label{fig:islands}
\end{figure}

\clearpage
The three island selection algorithms created Islands A, B, and C, respectively, shown in Figure~\ref{fig:islands}. Island A contains three substations: Winterpock substation, Chesterfield County substation, and Dinwiddie County substation. Two transmission lines must be disconnected to make Island A possible.
Island B contains 2 substations: the Rockville Substation and Short Pump Substation. Two transmission lines must be disconnected to make Island B possible.
Lastly, Island C contains only a single substation, the Iron Bridge Substation. Four transmission lines must be disconnected to make Island C possible. Candidates were rejected if isolating them fragmented the surrounding network, meaning that disconnecting Island C did not isolate any additional substations outside the island.

The search algorithms selected small groups of substations, limiting the number of switching actions needed to form the islands. Since protective islanding requires human labor and coordination, these small islands identify limited sets of actions for further evaluation, although their operational feasibility would still require checks of power flow and operating constraints.

\begin{figure}[!ht]
  \centering
  \includegraphics[width=\textwidth,keepaspectratio]{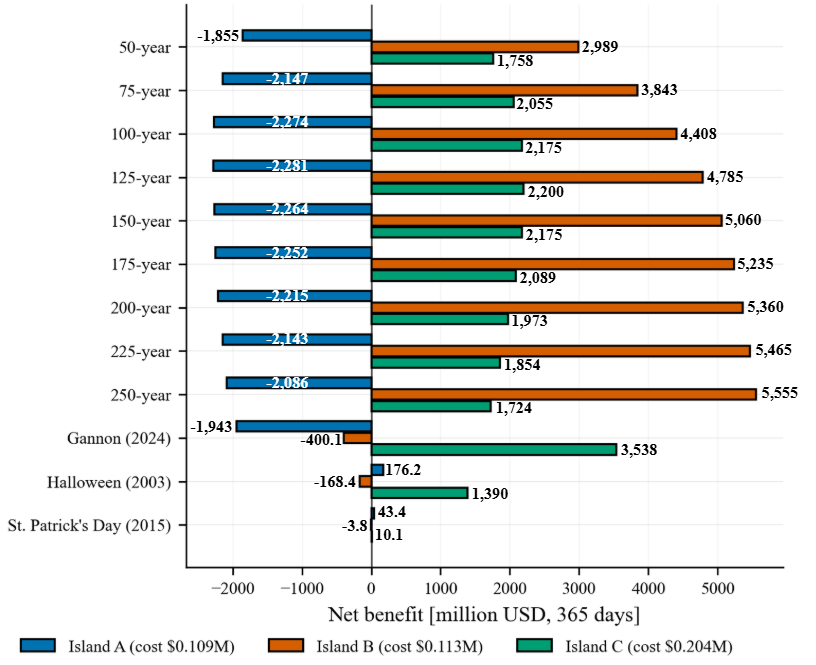}
  \caption{Net economic benefit by scenario for Islands A, B, and C using the Ghosh model, with transformer replacement after 365 days and disconnection for 6 hours.}
  \label{fig:islandbenefitcost}
\end{figure}

Costs and benefits were estimated for each island using the Ghosh input-output model, under the assumption of transformer replacement after 365 days and disconnection for 6 hours, as shown in Figure~\ref{fig:islandbenefitcost}. The one-time cost of implementing each island, comprising switching cost and the cost of the island's own temporary outage, was estimated at \$0.109M for Island A, \$0.113M for Island B, and \$0.204M for Island C. Notably, Island C was the most expensive of the three to implement, yet, as discussed below, it also produced the largest and most consistent net benefit across tested scenarios. Island A increased modeled economic losses in 10 out of 12 tested scenarios. For those scenarios, the estimated Ghosh losses after islanding exceeded the losses without islanding, consistent with adverse redistribution of GIC across the network. This supports the assertion that considering redistribution across the network following islanding is imperative. Isolating a portion of the power grid does not guarantee a reduction of GIC impact.

For Island B, estimated benefits exceeded costs across all nine return period scenarios, with benefit-cost ratios of tens of thousands to one. At the same time, benefit was negative for all tested historical storms, suggesting a disconnect between Island B's algorithmic selection criteria and the drivers of economic losses for the historical storms.

For Island C, estimated benefits exceeded costs across all twelve tested scenarios. The benefit-cost ratios were typically of a magnitude of several thousand to one, though the benefit-cost ratio was only $\approx50$ in the case of the 2015 St. Patrick's Day storm. Island C therefore produced a positive net benefit across the evaluated scenarios, although the three historical storms were also used to select the island.

\section{Conclusions and discussion}

This work presents the first adaptation at the state level of the U.S. Coupled Space Weather Impact Model (C-SWIM) presented in \citeA{oughton2026majorspaceweatherrisks}. By applying the methodology of the C-SWIM model to the state of Virginia, this work's contributions include a reproducible framework for regional adaptation of a national assessment that couples physics, engineering, and economics, an actionable assessment of geomagnetic storm risk for Virginia, and a method of protective islanding informed by data and evaluated through cost-benefit analysis.

The adaptation framework is a key transferable result of this work. 
Adaptation required considering factors that depend on topology and factors that are independent of topology to determine which components of the national assessment could be used for the regional network and which needed to be modified. This distinction provides a template for adapting the assessment to other subnational networks, with the network layout, connections, and input data checked for each application.

\emph{What are the specific power grid vulnerabilities for the state of Virginia under varying geomagnetic storm scenarios?}

Adapted to Virginia, the pipeline resolved geomagnetic risk to the level of individual substations, the level at which mitigation can occur. Risk was found to concentrate in central Virginia, particularly the Richmond region, where elevated estimated geoelectric fields coincide with a dense convergence of high-voltage transmission corridors, as shown in Figures~\ref{fig:transformerfailure} and \ref{fig:hazardmap}. The substation in Dinwiddie County at $37.25^{\circ}$ N, $77.61^{\circ}$ W had the highest mean simulated winding GIC for the 100-year scenario (186.4 A/ph mean, 320.9 A/ph 95th percentile), followed by North Pole Substation (109.8 A/ph mean, 289.2 A/ph 95th percentile).

\emph{How do power grid vulnerabilities translate into economic losses and individuals impacted for the state of Virginia?}

In measuring economic impact, total daily losses were estimated at \$145.5M for a 100-year storm and \$230.6M for a 250-year event, with between 339,000 and 535,000 people and up to 12,000 businesses at risk. The sectors with the largest estimated losses were manufacturing and finance, insurance, and real estate. These rankings of substations and regional loss estimates can inform transmission operators and state energy offices assessing mitigation in Virginia.

\emph{How can mitigation informed by data effectively be used to reduce the economic losses incurred due to geomagnetic storms?}

The exploration of protective islanding for mitigation produced the study's most instructive findings. The algorithm that produced Island A used betweenness centrality and a substation criticality index, and isolating Island A increased modeled economic losses in 10 out of 12 tested scenarios. Its selection criteria did not account for GIC redistribution outside the island. Island B's search algorithm incorporated redistribution analysis and current-flow betweenness, with an estimated Ghosh benefit of \$5.55B and a cost of \$0.113M for the 250-year scenario (benefit-cost ratio approximately 49,200), assuming transformer replacement after 365 days. However, Island B reduced estimated losses only for the return period scenarios. Under the historical storm scenarios, it increased losses, with a net economic penalty of up to \$400.1M for the Gannon storm. Island C, selected using solutions of the admittance matrix equations for the three historical storms, had benefits exceeding costs across all twelve evaluated scenarios, with benefit-cost ratios ranging from 50 for the St.~Patrick's Day storm to 17,300 for the Gannon storm over 365 days. The total benefit, though always positive for Island C, is highly variable across return periods. The sample standard deviation is $\approx\$790.4$M for all tested storms, but $\approx\$183.5$M excluding historical storms. The smaller spread describes variation among the nine fitted return period scenarios, which share the pattern of positive and negative line voltages derived from the Gannon storm. For the evaluated Virginia network and duration assumptions, protective islanding could reduce estimated economic losses when selection accounted for redistribution across the network and storm directionality. However, Island C was selected using the same historical storms included in its evaluation, so performance under other storms remains to be tested.

We applied a disruption localized to Virginia using national production and final demand tables, so the estimated economic consequences depend on national relationships between sectors. Transformer types and numbers of units at each substation were also randomly sampled because public equipment data were limited. The fragility calculation used one third of the absolute neutral current as a proxy for current per phase, which does not represent effective winding current for every transformer configuration. Moreover, transformer thermal failure was modeled and voltage collapse, cascading failure, and restoration were excluded. The estimated islanding benefits depend on the assumed durations of disconnection and transformer replacement. Future work includes constructing a social accounting matrix for the region, applying the adaptation framework to additional regions, and transitioning analysis from protective islanding to resilience islanding, which relies on local generation to maintain supply within a disconnected island.

\clearpage
\appendix
\section{Symbols for island selection}
\label{app:islanding-symbols}
\begingroup
\footnotesize
\begin{longtable}{c c p{9.5cm}}
\caption{Symbol definitions for the three islanding selection methods: A (SCI based on betweenness centrality), B (current-flow betweenness with redistribution measurement), and C (candidate identification using current-flow betweenness, with scoring based on admittance calculations for historical storms).}
\label{tab:islanding-legends}\\
\toprule
\rowcolor{tblhead}
\textbf{Method} & \textbf{Symbol} & \textbf{Definition} \\
\midrule
\endfirsthead
\multicolumn{3}{l}{\textit{Table~\ref{tab:islanding-legends} continued}} \\
\toprule
\rowcolor{tblhead}
\textbf{Method} & \textbf{Symbol} & \textbf{Definition} \\
\midrule
\endhead
\midrule
\multicolumn{3}{r}{\textit{continued on next page}} \\
\endfoot
\bottomrule
\endlastfoot
\multirow{9}{*}{A} & $G$ & Power grid network graph (substations as nodes, transmission lines as edges) \\
 & $b_i$ & Betweenness centrality of substation $i$ \\
 & $r_i$ & Normalized risk from geomagnetically induced current (GIC) at substation $i$ \\
 & $e_i$ & Normalized economic exposure of substation $i$ (population and GDP density) \\
 & $\text{SCI}_i$ & Composite substation criticality index, a weighted combination of $b_i$, $r_i$, $e_i$ \\
 & $S$ & Set of eligible substations (connected to the grid, within the study region) \\
 & $V$ & Total criticality value of the candidate island, $V = \sum_{i \in \text{island}} \text{SCI}_i$ \\
 & $C$ & Search cost score for boundary lines, summed as nominal voltage divided by 100~kV \\
 & $R$ & Ratio of criticality to cost, $R = V / C$ \\
\midrule
\multirow{10}{*}{B} & $G$ & Power grid network graph (substations as nodes, transmission lines as edges) \\
 & $r_i$ & Normalized risk from geomagnetically induced current (GIC) at substation $i$ \\
 & $e_i$ & Normalized economic exposure of substation $i$ (population and GDP density) \\
 & $w_i$ & Risk weight of substation $i$, $w_i = 0.5\,r_i + 0.5\,e_i$ \\
 & $\text{cfb}_i$ & Current-flow betweenness centrality of substation $i$, recomputed before and after disconnecting the candidate island's boundary lines \\
 & $\text{benefit}$ & Sum of $w_i$ over the candidate island's own members \\
 & $\text{redistribution\_cost}$ & Weighted change in current-flow betweenness at substations outside the island, caused by disconnecting the boundary lines \\
 & $\text{switching\_cost}$ & Search cost score for boundary lines, summed as nominal voltage divided by 100~kV \\
 & $\lambda,\ \mu$ & Weighting coefficients balancing redistribution cost and switching cost against benefit \\
 & $\text{net\_score}$ & $\text{benefit} - \lambda \cdot \text{redistribution\_cost} - \mu \cdot \text{switching\_cost}$ \\
\midrule
\multirow{9}{*}{C} & $G$ & Power grid network graph (substations as nodes, transmission lines as edges) \\
 & $w_i$ & Risk weight of substation $i$ (GIC risk and economic exposure), used only to help identify candidate islands \\
 & $\text{cfb}_i$ & Current-flow betweenness centrality of substation $i$, used only to identify more candidate islands, never to score a candidate \\
 & $\text{own\_effect}$ & Change in expected GDP loss at the candidate island's members, weighted by assumed failure probabilities for one historical storm \\
 & $\text{redistribution\_cost}$ & Change in expected GDP loss at substations outside the island, weighted by assumed failure probabilities for one historical storm \\
 & $\text{worst\_case}$ & The largest (most unfavorable) $\text{own\_effect} + \text{redistribution\_cost}$ across the three historical storms considered \\
 & $\varepsilon$ & Required safety margin: a candidate is only accepted if $\text{worst\_case} \le -\varepsilon$ \\
 & $\mu$ & Weighting coefficient applied to switching cost \\
 & $\text{net\_score}$ & $-\text{worst\_case} - \mu \cdot \text{switching\_cost}$, computed only for candidates that satisfy the required safety margin \\
\end{longtable}
\endgroup
\section*{Acknowledgments}\label{acknowledgements}
Dennies Bor and Edward Oughton gratefully acknowledge project funding via the 2026 UCAR PCEF Award entitled "Community-Driven Multi-Hazard Risk Assessment of Virginia: Quantifying Population and Built Environment Vulnerability".

\section*{Open Research}\label{open-research}
\addcontentsline{toc}{section}{Open Research}

Code for the Virginia adaptation and protective islanding analysis is available in the project repository (\url{https://github.com/VedaangV/c-swimvv}). The datasets \emph{C-SWIM Data} \cite{cswim_data_2026} and \emph{Coupled Space Weather Impact Modeling Data} \cite{cswim_data_2025} are available from Zenodo. The source data include geomagnetic observations from INTERMAGNET, the U.S. Geological Survey, and Natural Resources Canada, MT transfer functions from EarthScope \cite{Kelbert2011,Kelbert2017}, substation locations from OpenStreetMap, and transmission lines from HIFLD. Population and establishment data are available from the U.S. Census Bureau, economic data from the U.S. Bureau of Economic Analysis, and land cover data from the National Land Cover Database. The repository contains the corresponding data acquisition and processing scripts.

\section*{Conflict of Interest}
The authors declare there are no conflicts of interest for this manuscript.

\newpage
\bibliography{references}
\end{document}